\documentclass{webofc}
\usepackage[varg]{txfonts}   % Web of Conferences font
\begin{document}
\title{Mapping Datasets to Object Storage System}

\author{\firstname{Xiaowei(Aaron)} \lastname{Chu}\inst{1,3}\fnsep\thanks{\email{xweichu@ucsc.edu}} \and
        \firstname{Jeff} \lastname{LeFevre}\inst{2}\fnsep\thanks{\email{jlefevre@ucsc.edu}} \and
        \firstname{Aldrin} \lastname{Montana}\inst{3}\fnsep\thanks{\email{akmontan@ucsc.edu}}
        \firstname{Dana} \lastname{Robinson}\fnsep\thanks{\email{derobins@hdfgroup.org}}
        \firstname{Quincey} \lastname{Koziol}\fnsep\thanks{\email{koziol@lbl.gov}}
        \firstname{Peter} \lastname{Alvaro}\fnsep\thanks{\email{palvaro@ucsc.edu}}
        \firstname{Carlos} \lastname{Maltzahn}\fnsep\thanks{\email{carlosm@ucsc.edu}}
        % etc.
}

\institute{University of California, Santa Cruz}

\abstract{%
Access libraries such as ROOT\cite{rademakers1998root} and HDF5\cite{folk2011overview} allow users to interact with datasets using high level abstractions, like coordinate systems and associated slicing operations. Unfortunately, the implementations of access libraries are based on outdated assumptions about storage systems interfaces and are generally unable to fully benefit from modern fast storage devices. For example, access libraries often implement buffering and data layout that assume that large, single-threaded sequential access patterns are causing less overall latency than small parallel random access: while this is true for spinning media, it is not true for flash media. The situation is getting worse with rapidly evolving storage devices such as non-volatile memory and ever larger datasets. This project explores distributed dataset mapping infrastructures that can integrate and scale out existing access libraries using Ceph’s extensible object model, avoiding re-implementation or even modifications of these access libraries as much as possible. These programmable storage extensions coupled with our distributed dataset mapping techniques enable: 1) access library operations to be offloaded to storage system servers, 2) the independent evolution of access libraries and storage systems and 3) fully leveraging of the existing load balancing, elasticity, and failure management of distributed storage systems like Ceph. They also create more opportunities to conduct storage server-local optimizations specific to storage servers. For example, storage servers might include local key/value stores combined with chunk stores that require different optimizations than a local file system. As storage servers evolve to support new storage devices like non-volatile memory, these server-local optimizations can be implemented while minimizing disruptions to applications. We will report progress on the means by which distributed dataset mapping can be abstracted over particular access libraries, including access libraries for ROOT data, and how we address some of the challenges revolving around data partitioning and composability of access operations.
}
\maketitle
\section{Introduction}\label{sec:introduction}
\label{intro}
Data-intensive scientific application domains such as computational and comparative genomics, finance, astrophysics, high-energy physics, climate science, and oil \& gas are looking to file systems to lessen the burden of their data management overhead without having to replace tool chains that took years to establish and have become an essential component in many complex workflows. Most of these tools rely either directly on POSIX IO file system abstractions or indirectly via access libraries such as HDF5, NetCDF4\cite{rew1990netcdf}, or ROOT. While the POSIX IO file system abstraction of named byte streams using a hierarchical name space appears very simple, a general solution to the mapping of complex scientific datasets to files that is also performant has so far been elusive. One of the reasons is that the byte stream file abstraction removes all logical structure from the data so that it becomes difficult for a storage system to align its internal data structures to the logical structure of the data or take an active role in optimizing data access. Access libraries somewhat address this problem by mapping dataset abstractions such as multi-dimensional arrays to files. This “middleware” was very successful because it kept the (often proprietary) storage interface general while allowing scientists to access datasets on a high level of abstraction. But this success was only temporary because the middleware’s design often hard-wires assumptions about the underlying storage system in terms of storage devices, internal data structures like pages, access pattern performance profiles,  and striping strategies. These assumptions are increasingly inadequate as network and storage devices are becoming much faster and the storage hierarchy much deeper and more heterogeneous.

Important access libraries have started to address this problem by introducing a new layer just below its API that allows the introduction of new backends with more up-to-date assumptions while not requiring any changes to the application. For example, HDF5 has introduced the Virtual Object Layer (VOL)\cite{mu2019interfacing} that has enabled an ecosystem of plugins that map the HDF5 API not only to its traditional Virtual File Layer (effectively re-implementing an entire file system on top of file) but also to object-based storage backends, including Amazon S3 and Ceph/RADOS\cite{weil2006ceph}. Another example is NetCDF4 which maps to HDF5. We expect that this design pattern of separating the implementations of the access library API from the implementations of backends will be introduced in other access libraries as well, e.g. we know of ongoing work in the ROOT access library community.  

The proposed project relies on two key insights: The first is that new backend abstractions allow access libraries to map onto themselves: for example, an HDF5 VOL layer can map to an HDF5 API. The second key insight is that combined with programmable storage abstractions available in Ceph \cite{programmable}, this backend abstraction allows pushing native access library code into the object layer and distribute it over many servers. To stay with the HDF5 VOL example, a VOL plugin can map to storage objects, each of which offer the HDF5 API via an unmodified HDF access library and map it to a VOL plugin that interfaces with the local object storage layer. Thus, there are two kinds of plugins: one at the client that scatter/gathers requests to objects, and one that maps HDF5 requests to an object to the local object storage layer consisting of a local file system or a combination of a key/value and chunk store like Ceph’s BlueStore. Another example is SkyhookDM \cite{lefe2019} which is a Ceph plugin. SkyhookDM uses fast in-memory serialization libraries such as Google Flatbuffers\cite{flatbuffers}, Apache Arrow\cite{arrow:apacheblog16} to re-organize the data objects stored in Ceph. The custom object classes in Ceph enables SkyhookDM to conduct remote operations such as select, project, filter, aggregate and compress. The RocksDB system on each Ceph storage server is used to build the remote indexing system. Similar to the HDF5 VOL example, SkyhookDM has a python client library which accepts queries from clients, issues sub-queries to objects, and collects results and returns them to clients. 

\section{Goals}\label{sec:goals}
\begin{figure}[h]
\centering
\includegraphics[width=10cm,clip]{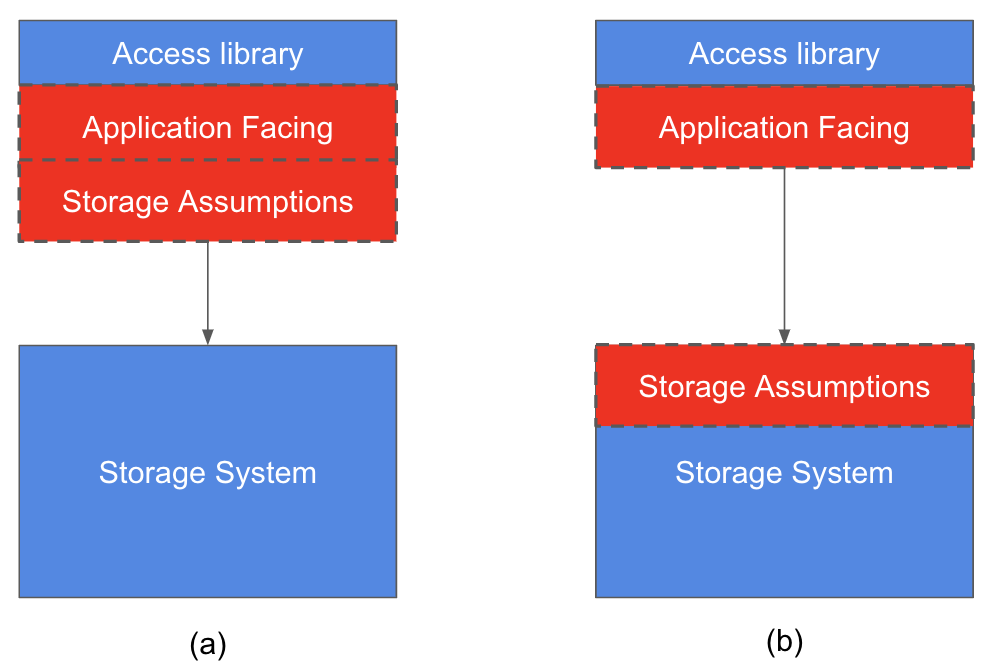}
\caption{Access Library Structures}
\label{fig-1}       % Give a unique label
\end{figure}

As Figure \ref{fig-1} (a) shows, the data access library can be split into two parts: one part is application facing which defines the public functions and interfaces. Applications use the functions and interfaces to access and manipulate the data models without worrying about the details of the storage system. The other part is storage system facing which is built based on the storage system assumptions. Most existing access libraries are built based on the file system interface and some special file formats are invented such as HDF5, ROOT, and NetCDF. Access libraries make basic assumptions about the storage system and organize the data bytes in the files accordingly. However, as we discussed in section \ref{intro}, such assumptions are relatively constant as the underlying storage systems changing rapidly. In addition, access libraries are usually designed for a single work station and do not scale out, thus handling datasets with large sizes is a big challenge. One potential solution is to re-implement the access libraries from time to time to accommodate the changes and utilize the novel features of the storage system. Obviously, this solution costs too much and the assumptions about the storage system will always be outdated. 
\\
The goals of this project are: (1) Make semantics of data available to the storage system so that it can manage the data more wisely. For example, the storage system can gather the data which is from the same logical units and put the data in the same storage locations such as objects or blocks. This can potentially reduce the number of the requests to the storage system and the amount of data needs to be moved when clients ask for the logical units. (2) Scale out the access library APIs and offload the operations to storage system. Assume that the storage system understands the semantics of the data and the logical units in the files are stored properly, operations such as ``\textsc{select}'', ``\textsc{project}'' , ``\textsc{aggregate}'' and ``compress'' can be offloaded to the storage system. This can greatly reduce the usage of CPU and memory resources from the clients side when analyzing datasets with large sizes by distributing the computations to the storage systems. For example, the Ceph distributed object storage system offers an object class extension feature. The object class feature allows users to effectively customize \texttt{read()} and \texttt{write()} methods for objects, and these custom methods can be used  for data processing and thus enable clients to push down some computations to run against the objects. SkyhookDM which is built on this feature is a good example to show off how database operations can be offloaded to Ceph storage system. (3) Allow the independent evolution of the access library and the backend storage systems. On one hand, the functions and interfaces of the access libraries rarely change as they are created according to the logical models of the data instead of storage system assumptions, thus re-implementation of the access libraries is not necessary. On the other hand, with the knowledge of the logical structures of the dataset, storage systems now have more flexibility to re-organize the data and conduct optimizations considering the characteristics of the local systems and devices without breaking the functionalities of the access libraries.

\section{Research Challenges} \label{sec:challenges}
 
The overall research goal of this project is to identify the means by which distributed dataset mapping infrastructures can be abstracted over particular access libraries while allowing those access libraries to evolve independently. Research challenges revolve around partitioning of the data, composability of access operations, and optimizations, both globally and locally as follows.

\subsection{Data Partitioning}
Data partitioning can have a great impact on the performance of storage systems since it has an impact on locality, parallel access, filtering, and load balancing. For example in a distributed storage system, if data is partitioned so that all input data for a common operation is on one server, that operation can be executed on that server without the need to transfer data. This is particularly important for holistic functions such as the median. For another example, object-based storage systems can store objects in a large range of sizes. The challenge is to find a size that both aligns with workload access patterns and strikes a good balance between parallel access and load balancing (smaller is better), and independent access and metadata overhead (larger is better).

\subsection{Composability of Access Operations}
Composability of access operations determines whether an access operation can be decomposed and distributed. Some operations are not de-composable and require that all input data needs to be transferred to the location where the operation is performed. The challenge is to reduce the amount of data transferred by either partitioning the data so that all input data is already at one location (see above) or by determining and using de-composable approximations that deliver acceptable results.

\subsection{Local and Global Optimizations}
Local and global optimizations can either complement or replace the ones implemented in access libraries. In the HDF5/VOL example, global optimizations can be performed by the client-side plugin (mapping VOL to HDF5 storage objects) while the plugin within HDF5 storage objects (mapping VOL to the local object storage backend) can perform local optimizations. Local optimizations will differ depending on the selection of available local storage systems such as a local file system, a local key/value store, or a local chunk store, and what storage devices these local storage systems operate on. In the proposed infrastructure, local optimizations can rely on information that is usually not available at traditional storage systems interfaces, such as on what tier(s) in the local storage hierarchy particular data is currently placed, or how access can be optimized for a particular storage device. As storage hierarchies and devices become more heterogeneous, local optimizers do not have to be known to clients and can be specialized to particular storage system configurations and deployments. Research challenges include how to automatically calibrate local optimizers when hardware, software, or configurations change, and how to communicate the capabilities of local optimizers to global optimizers in a sufficiently abstract way.

\section{Examples}\label{sec:examples}
\subsection{HDF5 VOL Example}
The Virtual Object Layer (VOL) is an intermediate layer in the HDF5 library sitting between the applications and the underlying storage systems. It intercepts all API calls that could access the data objects in an HDF5 file and forwards those calls to the VOL plugins. The plugins could store and organize the objects
in different ways based on the characteristics of the underlying storage systems. For example, a plugin could store the data objects in an object storage system and all the metadata to a database which speeds up the metadata lookups. The application still sees the same data model when operating on HDF5 files;
however, the VOL plugin is responsible to translate the original calls based on its own implementation which manages and accesses the data. The VOL abstraction leaves more freedom for the VOL plugins to optimize the data accesses for HDF5 files. It would also be possible to stack VOL plugins on top of each other when it is necessary. 
\begin{figure}[h]
\centering
\includegraphics[width=4.5cm,clip]{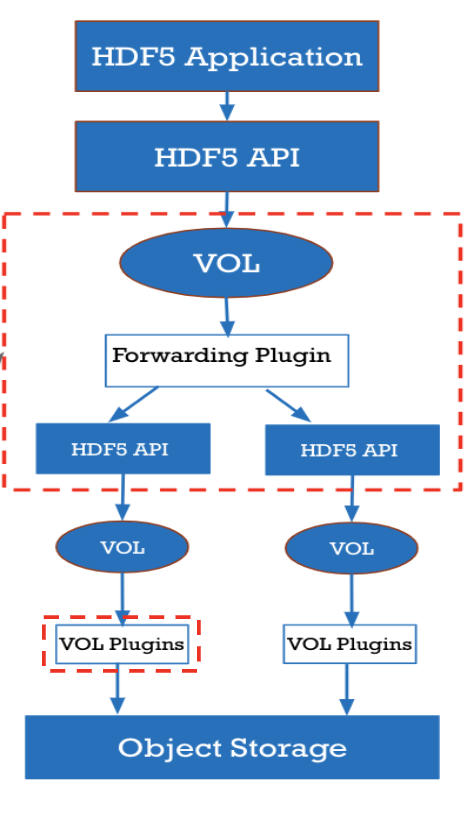}
\caption{HDF5 VOL Prototype Structure}
\label{fig-2}       % Give a unique label
\end{figure}
\\
Figure \ref{fig-2} shows the HDF5 VOL example structure. In this example, there are two VOL plugins. The global plugin and the object layer local plugins. The global plugin accepts the requests from the clients, decomposes the requests, and distributes the sub-requests to the objects. The object storage layer VOL plugin maps the data to its local storage system and executes the sub-request. As local object layer devices and system could vary case by case, the implementations of the VOL plugins can be customized to capture the local storage system characteristics.

Comparing to the original implementation of the HDF5 access library, this model introduces an extra forwarding plugin which also introduces additional overhead. The new model is not beneficial if the forwarding overhead of the "forwarding plugin" is dominant. However, enough parallelism could offset this overhead.  We have built a prototype system with multiple nodes to evaluate the performance of generating a 3GB dataset and try to figure out the parallelism needed to counteract this overhead. There are multiple nodes in the prototype system where HDF5 access library is installed on each node. The forwarding plugin "mirrors" the dataset writing requests and forwards them to one or more nodes. Table 1 shows the time it takes to create the same amount of data (3GB) by using different number of nodes in parallel.  
\begin{table}[]\label{tab1}
\centering
\begin{tabular}{|l|c|c|c|}
\hline
Number of Nodes & 1     & 2     & 3     \\ \hline
Time(s)         & 61.12 & 36.07 & 29.34 \\ \hline
\end{tabular}
\caption{Time it takes to create a 3GB dataset by different numbers of nodes in parallel}
\end{table}
For the same workload, it takes \textbf{26.28s} to finish by 1 node which uses the native HDF5 access library to write a 3GB dataset to one HDF5 file without the "forwarding plugin". And it takes 61.12s to finish the same workload for 1 node if there is the "forwarding plugin" involved. It takes 36.07s to finish by 2 nodes with each node writes 1.5GB dataset to a separate HDF5 file, the total size of the dataset is still 3GB. Similarly, 29.34s is taken for 3 nodes where each node writes 1GB dataset to a separate HDF5 file. For the last two cases, the "Forwarding Plugin" mirrors the dataset writing operations to multiple machines. According to data in the table, we can conclude that at least 3 nodes (29.34s is used to complete the workload) are required to work in parallel to offset the forwarding plugin overhead. 

\subsection{SkyhookDM Example}
SkyhookDM is built on the basis of the Ceph distributed storage system and its programmable storage feature. SkyhookDM and its python client enable the database operations to be offloaded to the storage system and shows the viability of our goals.

\begin{figure}[h]
\centering
\includegraphics[width=5cm,clip]{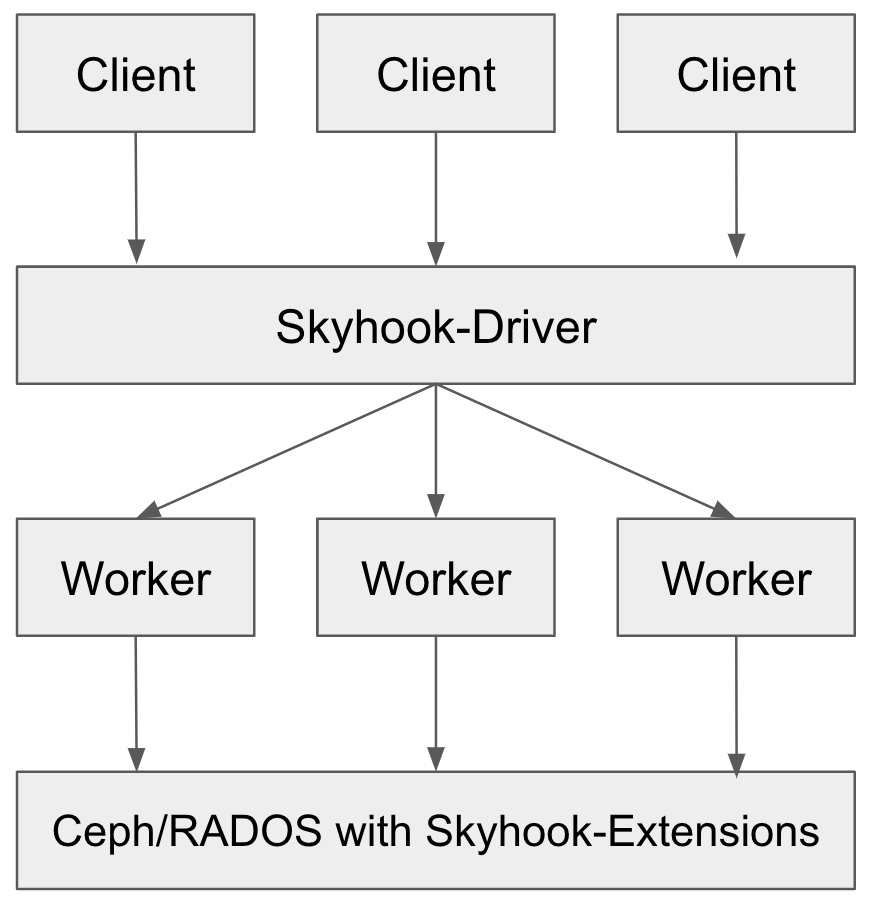}
\caption{SkyhookDM Design Architecture}
\label{fig-3}       % Give a unique label
\end{figure}

Tables are partitioned into a number of sub-tables which are serialized as Flatbuffers and stored as objects in the Ceph distributed storage system. Every node in the Ceph cluster has the Skyhook-Extension installed which is able to process the data in the objects.  Figure \ref{fig-3} shows the architecture of SkyhookDM with its python client.
\\
\begin{itemize}
            \item \textbf{Client:} submits data read, write and query requests to Skyhook-Driver. 

            \item \textbf{Skyhook-Driver:} schedules data writing and data processing tasks. It also aggregates the results returned by the Workers and passes the final outcomes to the clients. 

            \item \textbf{Worker}: for the data writing tasks, it partitions the data into objects, adds extra metadata and a format wrapper, and finally writes the objects to Ceph. For the data processing tasks, it operates on a set of objects and calls the Skyhook-Extension functions to process the data inside objects remotely, and further processes the data it gets from Skyhook-Extensions if necessary before it returns the results to Skyhook-Driver.   
            
            \item \textbf{Ceph/RADOS with Skyhook-Extensions:} stores and lays out the object data, creates indices which speed up data queries, and process the data inside the objects.
            
\end{itemize}

\begin{figure}[h]
\centering
\includegraphics[width=13.5cm,clip]{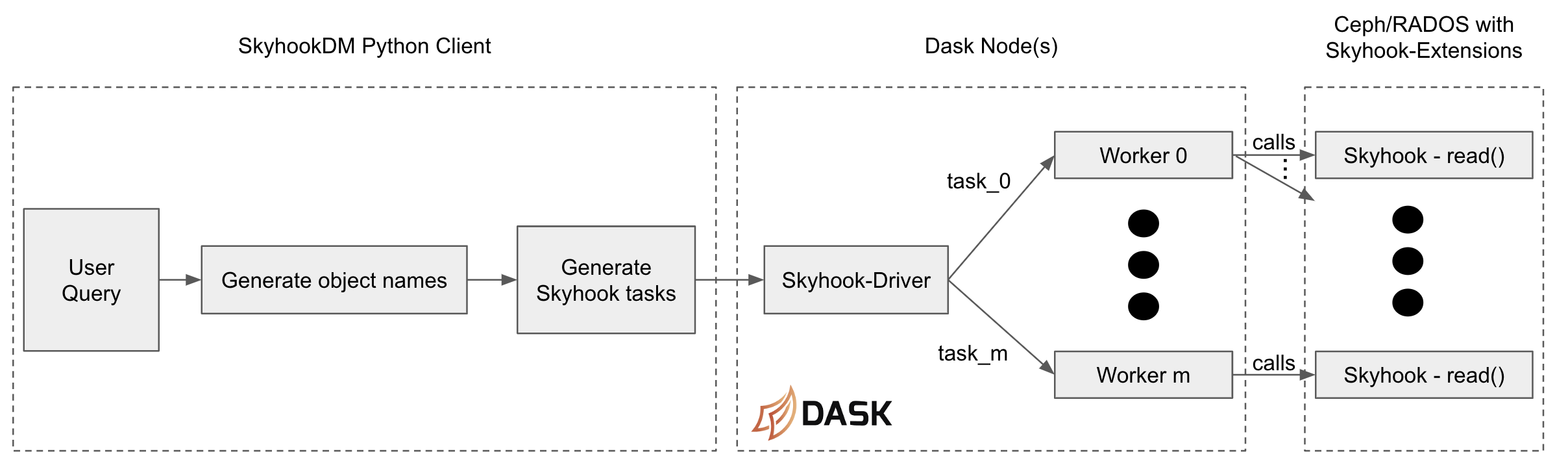}
\caption{SkyhookDM Workflow: Execute User Queries}
\label{fig-4}       % Give a unique label
\end{figure}

Dask\cite{rocklin2015dask} which is a scalable analytics framework in python is used to implement the system. Essentially, Skyhook-Driver is a Dask scheduler and Worker is a Dask worker. Both the Dask framework and the Ceph storage system with Skyhook have great scalability\cite{mascetti2015disk}, thus, components in the analytic workflow as shown in Figure~\ref{fig-4} can scale out independently to provide very flexible storage and data processing platforms. 

Figure \ref{fig-4} shows the workflow to execute a user query. Upon receiving the queries, the client generates object names and corresponding sub-queries, and send them to Skyhook-Driver. Skyhook-Driver schedules and dispatches the sub-queries to the worker threads. Sub-queries are further forwarded to Skyhook-Extensions to handle remotely at storage tier. Workers could further conduct some complicated computations against the results returned by Skyhook-Extensions. And finally, the outcomes are forwarded to the clients via Skyhook-Driver (Skyhook-Driver aggregates the results if necessary).  

In this example, data access libraries installed on the client side can evolve independently without making assumptions about the underlying storage systems. And Skyhook-Extensions potentially could have a number of customized implementations for each storage node. Comparing it to the HDF5 VOL example, Skyhook-Driver is similar to the role of the forwarding VOL plugin in the HDF5 VOL example, and Skyhook-Examples play the same roles as local VOL plugins in last example.

\section{Future Work}\label{sec:futurework}
Both the HDF5 VOL and SkyhookDM examples show the feasibility to resolve the current data management problems for scientific datasets. There are a lot interesting problems we're planning to investigate in future.
\begin{itemize}
    \item \textbf{Mapping datasets to objects of proper sizes.} Object storage systems usually have their own preferred object size ranges and some hard limitations on the max sizes of the objects. This requires: (1) identify the proper object size via experiments to make tradeoffs between parallelism and workload balance. (2) keep object size closer to the optimum size by grouping smaller logical units and further partitioning large logical units. (3) identify and co-locate relevant logical units when grouping. (4) Keep a minimum amount of metadata about the partition information.
    
    \item \textbf{Physical design management \cite{dahlgren2019towards}} Push down the data processing to the storage tier is challenging. Physical design management is required to optimize and speed up the data processing at the storage tier. For example, data transformation (row oriented to column oriented or vice versa) could be beneficial to some data query operations. This leads to problems such as striking for a balance between the cost of data transformation and workload performance improvement, online/offline data transformation, and local or distributed data transformation. Apart from the data transformation challenge, there are issues such as the management of indexing and local metadata remains to be resolved. 
    
    \item \textbf{Pushdown data processing operations.} As the datasets are broken down into objects, original data processing operations can not be executed directly any more. Existing access library functions and data processing operations might need to be transformed to sub-operations against the objects. In some scenarios, data transferring is inevitable, but we want to keep it minimum. In the meanwhile, strategies to schedule the sub-operations and recover from failures while executing are in our to-do list.
    
\end{itemize}

\section{Conclusion}\label{conclusion}
The implementations of existing access libraries are based on outdated storage system assumptions due to rapid development of the storage devices and network. In the meanwhile, access libraries are usually designed for a single workstation and can't scale out to handle extremely large dataset. Thus the applications can not fully benefit from the emerging technologies. We propose to let the storage system to understand the logical layout of the data so that operations can be offloaded to the storage systems and the characteristics of the storage system can be fully leveraged. This practice also allows the independent evolution of access libraries and the storage system. In addition, access libraries can be scaled out to handle large datasets in this way. We present the HDF5 VOL example and SkyhookDM example to show the viability of our method. And for the future work, we will focus on partitioning the datasets to objects, physical design managment, offloading data processing operations and some unforeseen difficulties and challenges.

\section{Acknowledgement}\label{sec:ack}
\label{ack}
This work was supported by NSF OAC 1836650 (IRIS-HEP), NSF CNS 1764102 (DeclStore), NSF CNS 1705021 (Programmable Storage), and Center for Research in Open Source Software (cross.ucsc.edu).

% For tables use syntax in table~\ref{tab-1}.
% \begin{table}
% \centering
% \caption{Please write your table caption here}
% \label{tab-1}       % Give a unique label
% % For LaTeX tables you can use
% \begin{tabular}{lll}
% \hline
% first & second & third  \\\hline
% number & number & number \\
% number & number & number \\\hline
% \end{tabular}
% % Or use
% \vspace*{5cm}  % with the correct table height
% \end{table}
%
% BibTeX or Biber users please use (the style is already called in the class, ensure that the "woc.bst" style is in your local directory)
% \bibliography{name or your bibliography database}

\bibliography{srl, chep}

\begin{thebibliography}{12}

\bibitem{rademakers1998root}
F.~Rademakers, R.~Brun et~al., Linux Journal \textbf{51}, 27 (1998)

\bibitem{folk2011overview}
M.~Folk, G.~Heber, Q.~Koziol, E.~Pourmal, D.~Robinson, \emph{An overview of the
  HDF5 technology suite and its applications}, in \emph{Proceedings of the
  EDBT/ICDT 2011 Workshop on Array Databases} (2011), pp. 36--47

\bibitem{rew1990netcdf}
R.~Rew, G.~Davis, IEEE computer graphics and applications \textbf{10}, 76
  (1990)

\bibitem{mu2019interfacing}
J.~Mu, J.~Soumagne, S.~Byna, Q.~Koziol, H.~Tang, R.~Warren, \emph{Interfacing
  HDF5 with A Scalable Object-centric Storage System on Hierarchical Storage},
  in \emph{2019 Cray User Group (CUG) meeting} (2019)

\bibitem{weil2006ceph}
S.A. Weil, S.A. Brandt, E.L. Miller, D.D. Long, C.~Maltzahn, \emph{Ceph: A
  scalable, high-performance distributed file system}, in \emph{Proceedings of
  the 7th symposium on Operating systems design and implementation} (2006), pp.
  307--320

\bibitem{programmable}
\emph{Dynamic object interfaces with lua},
  \url{https://ceph.com/rados/dynamic-object-interfaces-with-lua/}

\bibitem{lefe2019}
J.~LeFevre, N.~Watkins, \emph{Skyhook: Programmable Storage for Databases}, in
  \emph{Vault'19} ({USENIX} Association, Boston, MA, 2019)

\bibitem{flatbuffers}
\emph{{Google FlatBuffers} documentation},
  \url{https://google.github.io/flatbuffers/} (2015), accessed: 2019-08-01

\bibitem{arrow:apacheblog16}
{Apache Software Foundation} (2016), web Page.
  blogs.apache.org/foundation/entry/the\_apache\_software\_foundation\_announces87

\bibitem{rocklin2015dask}
M.~Rocklin, \emph{Dask: Parallel computation with blocked algorithms and task
  scheduling}, in \emph{Proceedings of the 14th python in science conference}
  (Citeseer, 2015), 130-136

\bibitem{mascetti2015disk}
L.~Mascetti, E.~Cano, B.~Chan, X.~Espinal, A.~Fiorot, H.G. Labrador, J.~Iven,
  M.~Lamanna, G.L. Presti, J.~Mo{\'s}cicki et~al., \emph{Disk storage at CERN},
  in \emph{Journal of Physics: Conference Series} (IOP Publishing, 2015), 4, p.
  042035

\bibitem{dahlgren2019towards}
K.~Dahlgren, J.~LeFevre, A.~Shirwadkar, K.~Iizawa, A.~Montana, P.~Alvaro,
  C.~Maltzahn, \emph{Towards Physical Design Management in Storage Systems}, in
  \emph{2019 IEEE/ACM Fourth International Parallel Data Systems Workshop
  (PDSW)} (IEEE, 2019), pp. 40--49

\end{thebibliography}

% \begin{thebibliography}{}
% % \bibliography{srl}
% %
% % and use \bibitem to create references.
% % Format for Journal Reference
% Journal Author, Journal \textbf{Volume}, page numbers (year)
% Book Author, \textit{Book title} (Publisher, place, year) page numbers
% \end{thebibliography}

\end{document}